\pdfoutput=1
\documentclass[aps,10pt,pre,byrevtex,twocolumn,bibnotes,footinbib,superscriptaddress,longbibliography]{revtex4-1}
\usepackage{microtype}
\usepackage{amsmath}
\usepackage{amssymb}
\usepackage{graphicx}
\usepackage{xcolor}
\usepackage{color}
\definecolor{myblue}{rgb}{0.153,0.322,0.706}
\usepackage[colorlinks,linkcolor=myblue,urlcolor=myblue,citecolor=myblue,breaklinks=true]{hyperref}

\newcommand{\be}{\begin{equation}}
\newcommand{\ee}{\end{equation}}
\newcommand{\ra}{\rightarrow}
\newcommand{\reals}{\mathbb{R}}
\newcommand{\tF}{\tilde F}
\newcommand{\tp}{\tilde p}

\newcommand{\tX}{\tilde X}

\newcommand{\cL}{\mathcal{L}}

\newcommand{\bF}{\bar F}
\newcommand{\bI}{\bar I}
\newcommand{\blambda}{\bar\lambda}
\newcommand{\bp}{\bar p}

\newcommand{\bA}{\bar A}
\newcommand{\bR}{\bar R}
\newcommand{\bL}{\bar L}

\begin{document}

\title{Linear approximations of large deviations: Cubic diffusion test}

\author{Pelerine Tsobgni Nyawo}
\affiliation{Department of Physics, University of Buea, Buea, Cameroon}

\author{Hugo Touchette}
\affiliation{Department of Mathematical Sciences, Stellenbosch University, Stellenbosch, South Africa}

\date{\today}

\begin{abstract}
We propose a method for approximating the large deviation rate function of time-integrated observables of diffusion processes, used in statistical physics to characterize the fluctuations of nonequilibrium systems. The method is based on linearizing the effective process associated with the large deviations of the process and observable considered, and is tested for a simple one-dimensional nonlinear diffusion model involving a cubic drift. The results show that the linear approximation compares well with the exact rate function, especially in the large fluctuation regime, and that its accuracy is related to the way the linearized process localizes in space. Possible extensions and applications to more complex diffusion models are proposed for future work.
\end{abstract}

\maketitle

\section{Introduction}

The theory of large deviations \cite{dembo1998,hollander2000,touchette2009,touchette2017,burenev2025} is commonly used to study the statistics of physical quantities related to nonequilibrium systems, such as particle and heat currents  \cite{derrida2007,bertini2007,hurtado2009b,harris2013}, the work exerted by external forces \cite{zon2003a,douarche2006,engel2009,sabhapandit2011} or the entropy production measuring the irreversibility of a process \cite{lebowitz1999,kurchan1998,harris2007,mehl2008}. These physical quantities or \emph{observables} play an important role in stochastic thermodynamics \cite{seifert2012,sekimoto2010,peliti2021} and are characterized in large deviation theory by the analog of a thermodynamic potential, called the \emph{rate function}, which determines the likelihood that an observable fluctuates around its typical value \cite{touchette2009,touchette2017,burenev2025}.

Many techniques have been developed over the years to compute rate functions by solving spectral and optimization problems related to large deviations, which are difficult to solve when considering processes involving many degrees of freedom or interacting particles. Alternatively, one can estimate rate functions by simulating trajectories using reweighting techniques, such as importance sampling and cloning \cite{giardina2006,lecomte2007a,angeli2019,kundu2011,nemoto2016,ferre2018,das2019,das2021b,singh2025,coghi2022}, which have the effect of biasing trajectories towards fluctuations of interest. This is an active area of research, currently, involving techniques from machine learning \cite{oakes2020,rose2021,causer2021,das2021,yan2022,gillman2024}.

In this paper, we propose a basic linearization method that can be used to approximate the rate function of diffusion processes, drawing on recent works on the large deviations of linear diffusion processes  \cite{buisson2022b}. The idea of the method is to approximate the so-called effective process associated with the large deviations of an observable by a linear diffusion, which is a solvable Gaussian process, and to express the rate function as the known control cost of this process, resulting in an upper bound on the true (a priori unknown) rate function. This is similar conceptually to using Gaussian models for approximating the solution of the Fokker--Planck equation or stochastic control problems for nonlinear systems \cite{roberts2003,kappen2005b,lu2017,martin2019}.

We test this approximation for a simple nonlinear diffusion process and observable in one dimension for which the exact rate function can be computed numerically by solving a quantum-like spectral problem. Comparing this rate function with the one obtained analytically by linearization, we show that the approximation provides good quantitative results, especially in the tails of the rate function describing large fluctuations, complementing other large deviation methods used to describe small fluctuations. From the results, we also propose a criterion based on the variance of the Gaussian process to predict the accuracy of the approximation.

The results are preliminary, but open the way for constructing useful approximations of rate functions for more general and realistic processes, combining potentially different methods that work in different fluctuation regimes. We propose possible extensions and applications in that direction for future work. Physically, the linear approximation is also useful, in that it shows that large fluctuations can be described in a simple way as being created by an effective Gaussian process involving only two parameters: a center (related to its mean) and a friction (related to its variance).

\section{Large deviation problem}

We consider systems modelled as Markov diffusion processes, evolving via the stochastic differential equation (SDE)
\be
dX_t = F(X_t) dt +\sigma dW_t,
\ee 
where $F(x)$ is the drift function, $W_t$ is a Brownian motion acting as a noise source, and $\sigma$ is the noise amplitude. Because the linear approximation will be illustrated for a one-dimensional process, we assume from the outset that $X_t\in\reals$ and $W_t\in\reals$, so $F(x)$ is a real function. The generalization of our results to diffusions in $\reals^d$ will be discussed at the end. We also assume that $F(x)$ is such that $X_t$ is ergodic, so it has a stationary density $p^*(x)$ solving the stationary Fokker--Planck equation \cite{risken1996}.

Given a trajectory $(X_t)_{0\leq t\leq T}$ of the system, we imagine measuring some observable $A_T$, which we take here to have the integrated form
\be
A_T=\frac{1}{T}\int_0^T f(X_t) dt,
\ee
where $f$ is an arbitrary real function that depends on the observable considered. The probability density $p_T(a)$ of this observable is difficult to obtain exactly, but is known in many cases to scale in the long-time limit according to 
\be
p_T(a) \approx e^{-TI(a)}.
\label{eqldt1}
\ee
This means that the limit
\be
I(a) = \lim_{T\ra\infty} -\frac{1}{T}\ln p_T(a)
\ee
exists, so that the dominant term of $p_T(a)$ is a decaying exponential with the observation time $T$. 

The rate of decay $I(a)$ is the \emph{rate function} that we want to calculate. This function is positive and generally has a unique zero for ergodic processes, located at the stationary value
\be
a^*=E_{p^*}[f(X)] = \int_\reals p^*(x)f(x)dx.
\ee 
The scaling in \eqref{eqldt1} thus predicts that $A_T$ concentrates on $a^*$ as $T\ra\infty$, so that $a^*$ is the typical value of $A_T$, and that fluctuations around this value are exponentially suppressed according to $I(a)$ as $T$ increases.

There are many methods used in large deviation theory to obtain the rate function. The most common are based on the calculation of the \emph{scaled cumulant generating function} (SCGF), defined by the limit
\be
\lambda(k) = \lim_{T\ra\infty} \frac{1}{T}\ln E[ e^{TkA_T}].
\ee
It is known that when this function is differentiable and strictly convex, $I(a)$ is the Legendre transform of $\lambda(k)$, meaning that
\be
I(a) = k(a) a-\lambda(k(a)),
\ee
where $k(a)$ is the unique root of $\lambda'(k)=a$. To obtain $\lambda(k)$, two methods can be followed:

1) \emph{Spectral method}: Compute the dominant eigenvalue of a linear differential operator $\cL_k$, called the tilted generator. The form of $\cL_k$ and the corresponding spectral problem that needs to be solved to find the SCGF is described in many works (see, e.g., \cite{touchette2009,touchette2017,burenev2025}), so we do not repeat them here. For our purposes, we note that the spectral method reduces for diffusions in $\reals$ to finding the quantum ground state of the potential
\be
V_k(x) = \frac{|U'(x)|^2}{2\sigma^2}-\frac{U''(x)}{2}-kf,
\label{eqqpot1}
\ee
where $U(x)$ is the potential associated with $F(x)$, that is, $F(x)=-U'(x)$. This is the method that we will use in Sec.~\ref{sectest} to benchmark the linear approximation. For more details on this method, including an explanation of the relation between $V_k$ and $\cL_k$, see \cite{touchette2017}.

2) \emph{Optimization method}: Solve the following optimization problem:
\be
\lambda(k) = \max_{\tF} \{k E_{\tp}[f(X)] -R(\tF||F)\},
\label{equncons1}
\ee
where
\be
R(\tF||F) = \frac{1}{2\sigma^2}E_{\tp}\big[|\tF(X)-F(X)|^2\big].
\label{eqcost1}
\ee
The maximization is over all drift function $\tF$ leading to a stationary distribution $\tp$ entering in the expectations above. Rather than solving the unconstrained maximization in \eqref{equncons1}, one can also solve the following constrained minimization to obtain the rate function directly:
\be
I(a) = \min_{\tF: E_{\tp}[f(X)]=a} R(\tF||F).
\label{eqcons1}
\ee
In this case, the minimization is over all drift $\tF$ such that the typical value of $A_T$ is changed to $a$ instead of $a^*$. 

The optimization method serves as the basis for the linear approximation presented in the next section. We refer to \cite{chetrite2015} for the derivation of the formula in \eqref{equncons1} and \eqref{eqcons1}, and for more information about their meaning. For this paper, it is important to note that the minimizing drift $\tF^a$ solving the constrained optimization in \eqref{eqcons1} is related to the maximizing drift $\tF_k$ solving the unconstrained optimization in \eqref{equncons1} by $\tF^a = \tF_{k(a)}$, $k(a)$ being the root entering in the Legendre transform. This follows because the unconstrained optimization is the Lagrange version of the constrained optimization \cite{chetrite2015}.  

The drift obtained in either case defines a new diffusion process $\tX_t$, called the \emph{effective, driven} or \emph{auxiliary process}, which differs from the original diffusion $X_t$ when $a\neq a^*$ or, equivalently, when $k\neq 0$ \cite{chetrite2013,chetrite2014,chetrite2015,jack2010b}. In terms of $k$, we write the SDE of the effective process as
\be
d\tX_t = \tF_k(\tX_t)dt+\sigma dW_t,
\ee
where $W_t$ is another Brownian motion perturbing the evolution of $\tX_t$. The interpretation of this process follows by considering all the trajectories of $X_t$ leading to a particular fluctuation $A_T=a$ away from $a^*$. In the limit $T\ra\infty$, it is known that this restricted set of trajectories can be described as a Markov process with the SDE above by choosing $k=k(a)$, so that $\tX_t$ can be seen as the process creating the fluctuation $A_T=a$. Knowing this process is thus important for understanding how fluctuations arise in many types of systems, including diffusion processes \cite{tsobgni2016,tsobgni2018,buisson2020b,mallmin2019}, random walks \cite{coghi2018b,gutierrez2021,gutierrez2021b}, interacting particles \cite{jack2019,dolezal2019,dolezal2022,ayyer2016,dagallier2025}, dynamical maps \cite{gutierrez2023,buca2019,wilkinson2022}, and active matter systems \cite{grandpre2018,das2022,fodor2020,keta2021,pineros2025}, among others.

\section{Linear approximation}

Finding the rate function or the SCGF is a challenging problem, in general, since it involves solving a spectral or optimization problem whose dimension grows with the number of degrees of freedom or particles considered. For this reason, numerical and approximation methods are often used to obtain these functions with a varying degree of accuracy. 

The approximation that we propose is based on assuming that the effective process is a linear diffusion characterised by the drift
\be
\bF(x) = -\alpha (x-\beta),
\label{eqFapprox1}
\ee
which involves two parameters: the friction $\alpha>0$ and the mode or center $\beta$, corresponding to the fixed point of the drift (in the noiseless dynamics). Since this class of drifts is a restriction on the set of all ergodic drifts, the constrained minimization in \eqref{eqcons1} applied to the parameters $(\alpha,\beta)$ must yield an upper bound on the true rate function. That is, if we define 
\be
\bI(a) = \min_{\alpha,\beta: E_{\bp}[f(X)]=a} R(\bF||F),
\label{eqans1}
\ee
then $\bI(a) \geq I(a)$. In this case, the minimization yields the best approximation to $I(a)$ that can be obtained with the linear ansatz, which is necessarily an upper bound to the rate function. The minimizing parameters $(\alpha^a,\beta^a)$ are the optimal parameters of that approximation. 

For the SCGF, we have
\be
\blambda(k) = \max_{\alpha,\beta} \{k E_{\bp}[f(X)] -R(\bF||F)\}\leq \lambda(k).
\label{eqans2}
\ee
In this case, the maximization yields the best lower bound on $\lambda(k)$ achievable with the linear approximation, characterised by optimal parameters $(\alpha_k,\beta_k)$ that now depend on $k$ rather than $a$. As in the general case, these parameters must be related to the those obtained by constraining on $A_T=a$  through the Legendre relation $\blambda'(k)=a$.

In general, it is known that the drift $\tF(x)$ of the effective process modifies the original drift $F(x)$ in a nonlinear way even if $F(x)$ is linear. However, in many cases, $\tF(x)$ is found to be locally linear around a displaced fixed point (see, e.g., \cite{tsobgni2016,tsobgni2018,buisson2020b}), so it is natural to approximate it as in \eqref{eqFapprox1}. We will see in the next section that this is not the same as linearizing $\tF(x)$, which would require some knowledge about the effective process, and thus about the SCGF and rate function. 

The choice of $\bF(x)$ assumes no such knowledge: it simply defines a Gaussian process leading after optimization to the approximation $\bI(a)$, which is expected to be close to $I(a)$ in cases where $\tF(x)$ itself is locally linear around some fixed point. This is studied in detail in the next section with a specific application. Naturally, if the modified drift $\tF(x)$ is linear, then the approximation is exact, so that $\bI(a) = I(a)$. This is known to happen when the original diffusion $X_t$ is linear and the observable is either a linear integral of $X_t$, a quadratic integral of $X_t$, or a linear contraction of the displacement $dX_t$ \cite{buisson2022b}.

The linear approximation follows Eyink's action principle \cite{alexander1997,eyink1996a,eyink1998}, which attempts to approximate the spectral large deviation problem using the Rayleigh--Ritz method by expanding the eigenfunction in some function space or basis. The difference here is that the expansion is carried out at the level of $\tF(x)$, which is approximated in the simplest way by $\bF(x)$. Choosing other function approximations for $\tF(x)$ \cite{das2019,das2021b,singh2025}, including neural networks \cite{yan2022}, lead in general to better approximations, which cannot however be treated analytically.

\section{Test case}
\label{sectest}

To test the linear approximation, we consider the following SDE for $X_t\in\reals$:
\be
dX_t = -\gamma X_t^3dt+\sigma dW_t,
\ee
where $\gamma$ and $\sigma$ are positive constants, together with the observable
\be
A_T = \frac{1}{T}\int_0^T X_t(X_t+1)dt.
\ee
The SCGF and rate function of this model cannot be found analytically, owing to the nonlinear drift and observable, but can be computed numerically, making it a good model to benchmark numerical and simulation methods \cite{nemoto2016,yan2022}. Its stationary distribution is given analytically by
\be
p^*(x) =\frac{1}{\Gamma(\frac{1}{4})}\left( \frac{2^3\gamma}{\sigma^2}\right)^{1/4}e^{-\gamma x^4/(2\sigma^2)},
\ee
following the fact that the drift can be derived from the quartic potential $U(x)=\gamma x^4/4$. From this result, we find the typical value of $A_T$ to be
\be
a^*=E_{p^*}[X(X+1)] = \sqrt{\frac{2\sigma^2}{\gamma}}\frac{\Gamma(\frac{3}{4})}{\Gamma(\frac{1}{4})}.
\label{eqastarcubic1}
\ee

\begin{figure*}[t]
\centering
\includegraphics[width=\textwidth]{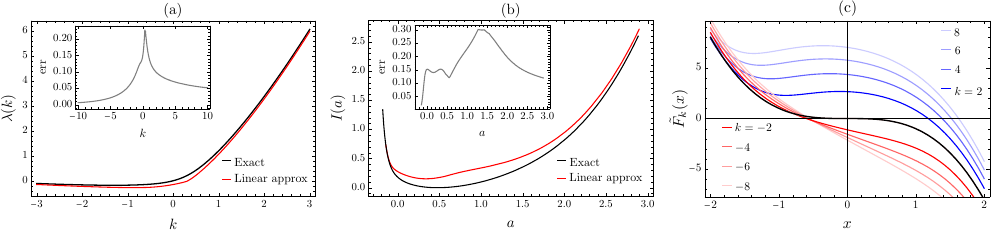}

\caption{(a) SCGF $\lambda(k)$ (black) compared with the linear approximation $\bar\lambda(k)$ (red). Inset: Error $\lambda(k)-\bar\lambda(k)$. (b) Rate function $I(a)$ (black) compared with the linear approximation $\bar I(a)$ (red). Inset: Error $\bar I(a)-I(a)$. (c) Effective drift $\tF_k(x)$ for different values of $k$ compared with the original drift $F(x)$ in black.}
\label{figld1}
\end{figure*}

The rate function describing the fluctuations of $A_T$ around this value is obtained with the spectral method mentioned before. First, we compute $\lambda(k)$ as the lowest eigenvalue of the quantum potential in \eqref{eqqpot1}, which leads with the quartic $U(x)$ above to a sextic potential \cite{kaushal1989}
\be
V_k(x) = \frac{\gamma^2 x^6}{2\sigma^2} -\frac{3\gamma x^2}{2}-kx(x+1).
\label{eqVk1}
\ee
To find the lowest eigenvalue of this potential, we used the NDEigensystem function of Mathematica with Dirichlet conditions, making sure to use a large-enough domain and fine-enough grid to ensure convergence. The result, yielding $\lambda(k)$, is shown in Fig.~\ref{figld1}(a) for $\gamma=\sigma=1$. Taking the Legendre transform then yields $I(a)$, shown in Fig.~\ref{figld1}(b) for the same parameters.

\begin{figure}[t]
\centering
\includegraphics{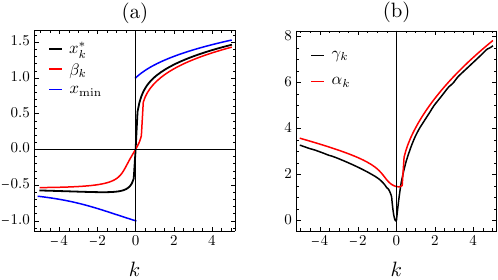}
\caption{Optimal parameters of the linear approximation $\bF(x)$ compared with the linearization of the effective drift $\tF_k(x)$. (a) Fixed point $x_k^*$ of $\tF_k(x)$ compared with the center $\beta_k$. The blue curve shows the minimum of $V_k(x)$. (b) Local friction $\gamma_k=-\tF'_k(x_k^*)$ compared with the friction $\alpha_k$.}
\label{figpar1}
\end{figure}

The results show two different fluctuation regions: one below the typical value $a^*\approx 0.478$, which is highly suppressed by a steep $I(a)$, and one above $a^*$ where $I(a)$ approximately grows quadratically with $a$, predicting Gaussian fluctuations. To understood how each region arises, we show in Fig.~\ref{figld1}(c) the drift of the effective diffusion, obtained from the ground-state wavefunction $\psi_k(x)$ associated with $\lambda(k)$ according to \footnote{This formula linking $\tF_k$ and $\psi_k$ applies to diffusions that can be symmetrised \cite{touchette2017}; see Sec.~C of \cite{angeletti2015} for a derivation.}
\be
\tF_k(x) = \sigma^2[\ln \psi_k(x)]'  =\sigma^2\frac{\psi_k'(x)}{\psi_k(x)}.
\ee
For $k>0$, corresponding to the fluctuations $a>a^*$, we can see that the original drift $F(x)=-\gamma x^3$ is modified to a nonlinear drift having a single fixed point $x_k^*$, satisfying $\tF_k(x_k^*)=0$, which increases with $k$. Moreover, the slope of $\tF_k(x)$ at that point decreases (negatively) with $k$, so the fluctuations $a>a^*$ can be seen as arising effectively from a diffusion having a modified center $x_k^*$ and local friction $\gamma_k=-F'_k(x_k^*)$  around $x_k^*$. A similar phenomenology applies to $k<0$, describing the fluctuations $a<a^*$. In that case, $x_k^*$ remains relatively constant with $k$ (see Fig.~\ref{figpar1}), while $\gamma_k$ increases as $k$ is decreased, explaining the different fluctuation regime.

The goal of the linear approximation is to capture these features of $\tF_k(x)$. Using the linear drift $\bF(x)$ in \eqref{eqFapprox1}, we find the optimal parameters $\alpha_k$ and $\beta_k$ that best approach $\tF_k(x)$ (without knowing this function) according to the maximization in \eqref{eqans2}. Since the stationary distribution associated with $\bF(x)$ is Gaussian,
\be
\bp(x) = \sqrt{\frac{\alpha}{\pi\sigma^2}}e^{-\alpha(x-\beta)^2/\sigma^2},
\ee 
the expectations involved in the maximization can be expressed analytically in terms of Gaussian moments. For the typical value of $A_T$ associated with a pair $(\alpha,\beta)$, we find
\be
\bA(\alpha,\beta) = E_{\bp}[X(X+1)] = \beta+\beta^2+\frac{\alpha}{2\sigma^2},
\ee 
while the control cost is
\begin{multline}
\bR(\alpha,\beta) =R(\bF||F) = \frac{15 \gamma ^2 \sigma ^4}{16 \alpha ^3}+\frac{45 \beta ^2 \gamma ^2 \sigma ^2}{8 \alpha ^2}+\frac{15 \beta ^4 \gamma ^2}{4 \alpha}\\
 -\frac{3 \gamma  \sigma ^2}{4 \alpha }+\frac{\alpha }{4}+\frac{\beta ^6 \gamma ^2}{2 \sigma ^2}-\frac{3 \beta ^2 \gamma }{2}.
\end{multline}
The cost to maximize in \eqref{eqans2} is therefore 
\begin{multline}
\bL(\alpha,\beta,k) = \displaystyle -\frac{15 \gamma ^2 \sigma ^4}{16 \alpha ^3}-\frac{45 \beta ^2 \gamma ^2 \sigma ^2}{8 \alpha ^2}-\frac{15 \beta ^4 \gamma ^2}{4 \alpha}+\frac{3 \gamma  \sigma ^2}{4 \alpha }\\
 \displaystyle -\frac{\alpha }{4}-\frac{\beta ^6 \gamma ^2}{2 \sigma ^2}+\frac{3 \beta ^2 \gamma }{2}+\frac{k\sigma ^2}{2 \alpha }+\beta ^2 k+\beta  k.
\end{multline}

Maximising this function for $\alpha>0$ and $\beta\in\reals$ is straightforward numerically. We show the results for the optimal $\alpha_k$ and $\beta_k$ as a function of $k$ in Fig.~\ref{figpar1}. The corresponding approximation $\blambda(k) = \bL(\alpha_k,\beta_k,k)$ is compared with the exact SCGF in Fig.~\ref{figld1}(a). 

As can be seen with the error plot shown in the inset, the linear ansatz gives a good approximation of $\lambda(k)$, despite its simplicity, especially in the tails. This can be understood by noting that the control cost $R(\tF||F)$ in \eqref{eqcost1} is an average, after optimization, of the distance between $\tF_k$ and $F$, weighted by the stationary distribution $p_k^*(x)$ of the effective process \footnote{The stationary distribution of the effective process is given by $p_k^*(x)=\psi_k(x)^2$, where $\psi_k(x)$ is the ground-state wavefunction associated with $\lambda(k)$ in the spectral problem~\cite{touchette2017}.}. Consequently, the region in space where $p_k^*(x)$ is largest contributes most to the cost, which corresponds here to the region around $x_k^*$ where $\tF_k(x)$ is locally linear, and thus where $p_k^*(x)$ is locally Gaussian. The more $p_k^*(x)$ is concentrated, the more this linear region contributes, leading to a better approximation by the linear ansatz.

This is confirmed in Fig.~\ref{figpar1}, which compares the optimal parameters of $\bF(x)$ with the linearization of $\tF_k(x)$. Overall, the linear ansatz captures the center and friction of $\tF_k(x)$ well, especially for large and small values of $k$. More crucial, however, is the fact that the friction of both drifts increases when $|k|$ increases, as seen in Fig.~\ref{figpar1}(b), resulting in a smaller error because of the concentration of $p_k^*(x)$  reproduced by $\bp(x)$.

\begin{figure}[t]
\centering
\includegraphics{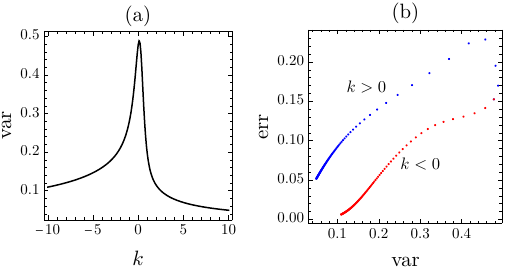}
\caption{(a) Variance of $p_k^*(x)$ as a function of $k$. (b) Relation between the error $\lambda(k)-\bar\lambda(k)$ and the variance, plotted parametrically in $k$.}
\label{figvar1}
\end{figure}

This is studied more quantitatively in the two plots of Fig.~\ref{figvar1}, which show, respectively, the variance of $p^*_k(x)$ as a function of $k$ and the error $\lambda(k)-\blambda(k)$ plotted parametrically with $k$ against the variance \footnote{The variance can be replaced with other quantities measuring the ``dispersion'' of $p_k^*(x)$, such as $1/\alpha_k$ or the inverse participation ratio of $\psi_k(x)$, with similar results.}. The latter plot shows two branches, coming from the two error tails around $k=0$, that confirm that the error decays with the variance. The decay is more pronounced for $k<0$ because the linear approximation is better there, as seen in Fig.~\ref{figld1}(a). 

Similar results apply to the rate function and are explained in the same way. In this case, the approximation $\bI(a) = \bR(\alpha_k,\beta_k)$ obtained for $a=\bA(\alpha_k,\beta_k)$ is good in the tails of $I(a)$, as shown in the error inset of Fig.~\ref{figld1}(b), especially for $a<a^*$. One interesting observation here is that $\bI(a)$ has an upward elevation in the middle, so it is nonconvex as a function of $a$. This is a feature of the linear approximation, associated with the discontinuity of $\alpha_k$ and $\beta_k$ that is visible in Fig.~\ref{figpar1} around $k=0.5$. The rate function itself is strictly convex in $a$, resulting in a SCGF that is differentiable. Moreover, $\tF_k(x)$ is continuous in $k$, so no discontinuity that could be interpreted as a dynamical phase transition appears in the model.

These results for the SCGF and rate function are qualitatively the same if we change $\gamma$ and $\sigma$ \footnote{The important parameter of the model is $\gamma/\sigma^2$, as can be seen from the form of $p^*(x)$.}. The error for both increases overall when $\sigma$ is increased or when $\gamma$ is decreased (high noise regime) and decreases otherwise for smaller values of $\sigma$ or larger values of $\gamma$ (low noise regime). This is expected, as the width of $p_k^*(x)$ increases with the noise level.

\section{Accuracy assessment}

The results described in the previous section are based on the exact (numerical) solution that we have for that particular model. What can be said about the accuracy of the linear approximation when the solution is not known? Can the approximation be used in a meaningful way to compute the SCGF and rate function?

Because the control cost $R(\tF_k||F)$ is a weighted expectation relative to the effective process, its drift need not be known or calculated accurately over the whole space; what matters most for building a good approximation of $\lambda(k)$ or $I(a)$ is to obtain $\tF_k(x)$ in the region where $p_k^*(x)$ is largest, which concentrates at low noise around the fixed points of the drift. With this insight and the results of the previous section, it is natural to conjecture that the linear approximation is accurate when: 

(i) $\tF_k(x)$ has a unique fixed point $x_k^*$, implying that $p_k^*(x)$ is unimodal;

(ii) $\tF_k(x)$ is locally linear around $x_k^*$, implying that $p_k^*(x)$ is locally Gaussian around that point; 

(iii) $p_k^*(x)$ is relatively concentrated around $x_k^*$. 

Under these conditions, the linear ansatz should capture the essential, linear features of $\tF_k(x)$ around $x_k^*$, as found for the cubic diffusion, leading to good approximations of the SCGF and rate function.

These conditions are fairly general and can be tested without knowing the effective process. We know, for instance, that this process is unimodal for a large class of observables when the original process itself is unimodal, so condition (i) above can be accepted on a general basis, at least in the absence of dynamical phase transitions. The same applies to condition (ii), which is a natural linearization assumption. Finally, condition (iii) can be tested with the knowledge of $\bp(x)$, which is found independently of $p_k^*(x)$. If the latter is locally Gaussian and has a small variance, so will $\bp(x)$, leading to a good approximation of $\lambda(k)$ and $I(a)$.

\section{Conclusion}

The results that we have presented for the simple cubic diffusion process show that the linear approximation or ansatz can be a useful tool for studying the large deviations of diffusion processes. Though quite basic, it can provide an accurate description of the effective process in the space mostly occupied by this process, which is important for determining the SCGF and rate function. For this reason, it should be used as a first approximation for quickly gaining information about these functions. In the worst case, bounds are obtained, which can be improved in principle by adding higher-order terms in the ansatz. From a numerical point of view, one can also use the linear ansatz as an initial guess in iteration schemes that solve the large deviation optimization problem \cite{ferre2018,yan2022}, leading potentially to faster convergence.

Future work should look at applications for higher-dimensional diffusions evolving in $\reals^d$ or bounded regions of $\reals^d$ with specified boundary conditions, such as reflective walls. For these, $x$ is a vector in $\reals^d$, the parameter $\alpha$ in $\bF(x)$ is a matrix, which has to be positive definite to ensure that the corresponding diffusion has a stationary distribution, and $\beta$ is a vector in $\reals^d$. For an application involving  a diffusion in $\reals$ with reflective boundaries, for which accurate approximations of the SCGF and rate function are also found, we refer to \cite{buisson2020b}.

Other types of linear approximations can be considered as replacement or complement to the one proposed here. Instead of assuming that $\tF_k(x)$ is linear, for instance, one could assume that it is a linear perturbation of the original drift $F(x)$, so that
\be
\bF(x) = F(x) -\alpha(x-\beta).
\ee
This is expected to give a good approximation of $\lambda(k)$ near $k=0$, and thus a good approximation of $I(a)$ near the typical value $a^*$. Using this ansatz leads to a simple expression of $|\bF(x)-F(x)|^2$ in the control cost \eqref{eqcost1}, but now $\bp(x)$ is not known analytically, in general, because $\bF(x)$ involves $F(x)$. In this case, one can rely on numerical methods to compute the cost, including Monte Carlo methods that estimate it over simulated paths.

Another natural approximation consists in replacing the quantum potential $V_k(x)$ in \eqref{eqqpot1} by a quadratic potential around its minimum. This harmonic approximation of the spectral problem also yields a linear form for the effective drift, which is different in principle from the one built on $\bF(x)$, although it may approach that drift in certain systems or fluctuation regimes. This happens, for example, for the cubic diffusion. The blue curve in Fig.~\ref{figpar1} shows the minima of $V_k(x)$ as a function of $k$, which is close to the center of $\bF(x)$ for large $|k|$, but not around $k=0$, so the harmonic approximation differs from the linear ansatz in that region. 

Beyond this application, it is important to note that the harmonic approximation can only be applied when the large deviation spectral problem can be ``symmetrized'' to a quantum problem, limiting this approximation to equilibrium diffusions, essentially \cite{touchette2017}. The linear ansatz is not limited in this way, since it is based on the optimization method. This means that it can be applied to nonequilibrium diffusions, including diffusions in $\reals^d$ with non-gradient drifts.

\section*{Acknowledgments}

We thank Dani\"el Cloete for comments on the paper.  P.T.N. is grateful to the Simons Foundation and the European Mathematical Society for a travel grant (EMS-Simons for Africa) that enabled her visit to Stellenbosch University.

\bibliography{masterbib}

\end{document}